\documentclass[aps,prl,twocolumn,superscriptaddress,showpacs,amsmath,amssymb]{revtex4}

\usepackage{graphicx}

\begin{document}


{\bf Comment on ``Coherent Ratchets in Driven Bose-Einstein Condensates''}\\

Creffield and Sols (henceforth CS) \cite{Creffield} recently reported
a finite, directed time-averaged ratchet current, for  \textit{noninteracting} quantum particles in a potential $V (x, t) = KV (x)f (t)$ with time-periodic driving $f (t)=f(t+T)$,
even when time-reversal symmetry holds, as depicted with the solid line in Fig. 3 in \cite{Creffield}.
CS chose $V (x) = \sin(x) + \alpha\sin(2x)$, $f (t) = \sin(t) + \beta \sin(2t)$ ($\beta=0$ in their Fig. 3) and the initial condition $\Psi(x, 0) = 1/\sqrt{2\pi}$.
As we will explain in the following, this result is incorrect, that is,
time-reversal symmetry implies a vanishing ratchet current.

The asymptotic time averaged current (TAC) is given by ${\displaystyle J=\lim_{\tau\rightarrow \infty} J(\tau)}$
where $J(\tau)=\tau^{-1}\int_0^{\tau}I(t)dt$. $I(t)$ is given
by $I(t)=-i \int_{-\infty}^{\infty}dx\;\Psi^*(x,t)\frac{\partial \Psi(x,t)}{\partial x}$.
Given the periodicity of the driving, $f(t)=f(t+T)$, one may analyze
the evolution in terms of the system's Floquet states. The asymptotic TAC is then given
by ~\cite{Denisov}
\begin{equation}
{\displaystyle J=\sum_{\alpha}
|C_{\alpha}|^2\left\langle \langle\psi_{\alpha}|\hat{p}|\psi_{\alpha}\rangle \right\rangle_T=\sum_{\alpha}
|C_{\alpha}|^2\left\langle \upsilon_{\alpha}(t) \right\rangle_T},
\end{equation}
where $\psi_{\alpha}$ are the Floquet eigen-states (FES),
$\psi_{\alpha}(t+T)=\psi_{\alpha}(t)$, the coefficients $C_\alpha$
are such that $\Psi(x,0)=\sum_{\alpha}C_{\alpha}\psi_{\alpha}(x,0)$,
$\upsilon_{\alpha}(t)=-i
\int_{-\infty}^{\infty}dx\;\psi_{\alpha}^*(x,t)\frac{\partial
\psi_{\alpha}(x,t)}{\partial x}$ is the instantaneous velocity of
the Floquet state,  and $\langle ... \rangle_T$ denotes the average
in time over the period $T$. The TAC for each FES vanishes
identically if $f(t_s+t)=f(t_s-t)$ for some $t_s$, because
$\upsilon_{\alpha}(t_s+t)=-\upsilon_{\alpha}(t_s-t)$, and therefore
$\langle \upsilon_{\alpha}(t) \rangle_T = 0$ \cite{Denisov}.  Given
that $J$ is the weighted sum (1), it follows that $J=0$ for
$\beta=0$ because $\sin(\pi/2 + t)=\sin(\pi/2-t)$. Since the
parameter $K$ does not change the symmetries of the system, and
given that the time-reversal symmetry implies a vanishing TAC, we
conclude that no asymptotic directed transport occurs for any value
of this parameter.
CS used the stroboscopic current,
${\displaystyle J_s(t_p,m)= \frac 1 {m+1} \sum_{n=0}^m I(t_p+nT)}$. Their asymptotic stroboscopic current is given
by \cite{Denisov}
\begin{equation}
{\displaystyle J_s(t_p)= \sum_{\alpha}
|C_{\alpha}|^2 \upsilon_{\alpha}(t_p)}= J_s(t_p+ T),
\end{equation}
where $\upsilon_{\alpha}(t)$ are periodic functions,
$\upsilon_{\alpha}(t+T)=\upsilon_{\alpha}(t)$. Since even in the
case of time-reversal symmetry instantaneous velocities are nonzero,
$\upsilon_{\alpha}(t_p)\neq 0$, the current (2) acquires a nonzero
value, which depends on the arbitrary choice of the measurement time
$t_p \in [0, T)$.

Motion is a continuous process and attempts to describe it in terms
of stroboscopic characteristics only may lead to wrong physical
conclusions. The harmonic oscillator constitutes a good example: Its
particle velocity is $\upsilon(t)=\upsilon_0 \sin[\omega (t - t_p)]$
and, depending on  $t_p$, the asymptotic stroboscopic averaged
velocity $v_s(t_p)$ may take any value  within the interval
$[-\upsilon_0, \upsilon_0]$, although no directed transport occurs.
\begin{figure}
\center
\includegraphics[width=0.38\textwidth]{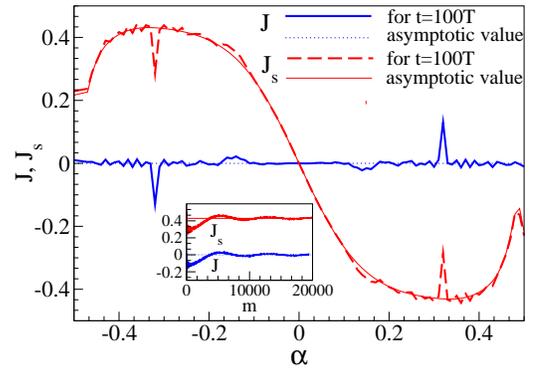}
\caption{(Color online) $J(t)$  and the stroboscopic current $J(0,m)$ as  functions of $\alpha$: for  $t=mT$, where $m=100$, thick blue solid line and thick red dashed line correspondingly; and their asymptotic values, $J$, Eq. (1), (thin blue dotted line) and $J_s$, Eq. (2)(thin red solid line).
Here $K=2.4$ and $\beta = 0$.
Inset:
Dependence of  $J(t=mT)$ (blue line) and $J_s(t_p=0, m)$ (red line) on $m$ at $\alpha = -0.32$. The thin lines are given by (1) and (2), respectively.} \label{fig1}
\end{figure}
We numerically verified the above conclusions
by performing an integration of the
Schr\"odinger equation with the same parameters as in
Fig. 3 of \cite{Creffield}. We used two independent methods \cite{Denisov, Poletti}.
The so obtained results do coincide and are depicted in our Fig. 1. For $\beta = 0$ we
numerically obtain virtually zero current for all values of $\alpha$, the thick (blue) solid line.
The amplitude of small fluctuations away from zero decrease systematically upon increasing
the overall integration time $\tau$, see  inset in Fig. 1.
These findings are therefore in full agrement with the symmetry analysis \cite{Denisov}.
In the contrast, the stroboscopic current used in  Ref. \cite{Creffield} remains finite forever, approaching values predicted by (2).
Moreover, the above symmetry analysis is not in
contrast with Ref. [3], where the  atom-atom
interactions obey time reversal symmetry.

G. Benenti$^1$,
G. Casati$^{1,2}$,
S. Denisov$^3$,
S. Flach$^4$,
P. H\"anggi$^{3,5}$,
B. Li$^5$, and
D. Poletti$^2$\\
 $^{1}$ CNR-INFM, Universit\`a degli Studi dell'Insubria, Via Valleggio 11, 22100 Como, Italy \\
 $^{2}$ Centre for Quantum Technologies, National University of Singapore, Singapore \\
 $^{3}$Institut f\"ur Physik, Universit\"at Augsburg, Universit\"atsstr. 1, 86135 Augsburg, Germany \\
$^{4}$Max Planck Institute for the Physics of Complex Systems, N\"othnitzer Str. 38, 01187 Dresden, Germany \\
 $^{5}$Department Physics and CSE, National University of Singapore, 117542 Singapore \\

\end{document}